\documentclass[10pt,journal,letterpaper,compsoc]{IEEEtran}

\usepackage{latexsym}
\usepackage{graphicx}
\usepackage{tabularx}
\usepackage{multirow}
\usepackage{color}
\usepackage{epsfig}
\usepackage{rotating}
\usepackage{amsmath}
\usepackage{dashrule}
\usepackage{fancyhdr}

\usepackage{hhline}
\usepackage{booktabs}
\usepackage{array}

\usepackage{afterpage}
\usepackage{pdflscape}
\usepackage{hyphenat}
\usepackage[table]{xcolor}
\usepackage{framed}
\usepackage{scrextend}
\usepackage{soul}

\usepackage{enumitem}

 \usepackage[bookmarks=true%
,bookmarksnumbered=true%
,hypertexnames=false%
,breaklinks=true%
,colorlinks=true%
,linkcolor=blue%
,citecolor=blue%
,urlcolor=blue%
,pdfpagelabels=false
]{hyperref}

\ifCLASSOPTIONcompsoc
\else
\fi

\ifCLASSINFOpdf
\else
\fi

\ifCLASSOPTIONcompsoc
\usepackage[tight,normalsize,sf,SF]{subfigure}
\else
\usepackage[tight,footnotesize]{subfigure}
\fi

\begin{document}

\title{Towards Real-Time, Country-Level Location Classification of Worldwide Tweets}

\author{
  Arkaitz Zubiaga$^1$, Alex Voss$^2$, Rob Procter$^1$, Maria Liakata$^1$, Bo Wang$^1$, Adam Tsakalidis$^1$ \\
  $^1$ University of Warwick, Coventry, UK \\
  $^2$ University of St Andrews, St Andrews, UK \\
  {\tt a.zubiaga@warwick.ac.uk}
}

\IEEEcompsoctitleabstractindextext{%
\begin{abstract}
 The increase of interest in using social media as a source for research has motivated tackling the challenge of automatically geolocating tweets, given the lack of explicit location information in the majority of tweets. In contrast to much previous work that has focused on location classification of tweets restricted to a specific country, here we undertake the task in a broader context by classifying global tweets at the country level, which is so far unexplored in a real-time scenario. We analyse the extent to which a tweet's country of origin can be determined by making use of eight tweet-inherent features for classification. Furthermore, we use two datasets, collected a year apart from each other, to analyse the extent to which a model trained from historical tweets can still be leveraged for classification of new tweets. With classification experiments on all 217 countries in our datasets, as well as on the top 25 countries, we offer some insights into the best use of tweet-inherent features for an accurate country-level classification of tweets. We find that the use of a single feature, such as the use of tweet content alone -- the most widely used feature in previous work -- leaves much to be desired. Choosing an appropriate combination of both tweet content and metadata can actually lead to substantial improvements of between 20\% and 50\%. We observe that tweet content, the user's self-reported location and the user's real name, all of which are inherent in a tweet and available in a real-time scenario, are particularly useful to determine the country of origin. We also experiment on the applicability of a model trained on historical tweets to classify new tweets, finding that the choice of a particular combination of features whose utility does not fade over time can actually lead to comparable performance, avoiding the need to retrain. However, the difficulty of achieving accurate classification increases slightly for countries with multiple commonalities, especially for English and Spanish speaking countries.
\end{abstract}

}

\maketitle

\IEEEdisplaynotcompsoctitleabstractindextext
\IEEEpeerreviewmaketitle

\section{Introduction}
\label{introduction}

Social media are increasingly being used in the scientific community as a key source of data to help understand diverse natural and social phenomena, and this has prompted the development of a wide range of computational data mining tools that can extract knowledge from social media for both post-hoc and real time analysis. Thanks to the availability of a public API that enables the cost-free collection of a significant amount of data, Twitter has become a leading data source for such studies \cite{weller2013twitter}. Having Twitter as a new kind of data source, researchers have looked into the development of tools for real-time trend analytics \cite{madani2015real,zubiaga2015real} or early detection of newsworthy events \cite{sakaki2010earthquake}, as well as into analytical approaches for understanding the sentiment expressed by users towards a target \cite{jiang2011target,kouloumpis2011twitter,townsend2015warwickdcs}, or public opinion on a specific topic \cite{bollen2011modeling}. However, Twitter data lacks reliable demographic details that would enable a representative sample of users to be collected and/or a focus on a specific user subgroup \cite{mislove2011understanding}, or other specific applications such as helping establish the trustworthiness of information posted \cite{middleton2016geoparsing}. Automated inference of social media demographics would be useful, among others, to broaden demographically aware social media analyses that are conducted through surveys \cite{duggan2015demographics}. One of the missing demographic details is a user's country of origin, which we study here. The only option then for the researcher is to try to infer such demographic characteristics before attempting the intended analysis.

This has motivated a growing body of research in recent years looking at different ways of determining automatically the user's country of origin and/or -- as a proxy for the former -- the location from which tweets have been posted \cite{ajao2015survey}. Most of the previous research in inferring tweet geolocation has classified tweets by location within a limited geographical area or country; these cannot be applied directly to an unfiltered stream where tweets from any location or country will be observed. The few cases that have dealt with a global collection of tweets have used an extensive set of features that cannot realistically be extracted in a real-time, streaming context (e.g., user tweeting history or social networks) \cite{dredze2016geolocation}, and have been limited to a selected set of global cities as well as to English tweets. This means they use ground truth labels to pre-filter tweets originating from other regions and/or written in languages other than English. The classifier built on this pre-filtered dataset may not be applicable to a Twitter stream where every tweet needs to be geolocated. An ability to classify tweets by location in real-time is crucial for applications exploiting social media updates as social sensors that enable tracking topics and learning about location-specific trending topics, emerging events and breaking news. Specific applications of a real-time, country-level tweet geolocation system include country-specific trending topic detection or tracking sentiment towards a topic broken down by country. To the best of our knowledge, our work is the first to deal with global tweets in any language, using only those features present within the content of a tweet and its associated metadata. We also complement previous work by investigating the extent to which a classifier trained on historical tweets can be used effectively on newly harvested tweets.

Motivated by the need to develop an application to identify the trending topics within a specific country\footnote{http://www.bbc.co.uk/programmes/b04p59vr}, here we document the development of a classifier that can geolocate tweets by country of origin in real-time. Given that within this scenario it is not feasible to collect additional data to that readily available from the Twitter stream \cite{dredze2016geolocation}, we explore the usefulness of eight tweet-inherent features, all of which are readily available from a tweet object as retrieved from the Twitter API, for determining its geolocation. We perform classification using each of the features alone, but also in feature combinations. We explore the ability to perform the classification on as many as 217 countries, or in a reduced subset of the top 25 countries, as judged by tweet volume. The use of two datasets, collected in October 2014 and October 2015, gives additional insight into whether historical Twitter data can be used to classify new instances of tweets. These two datasets with over 5 million country-coded tweets are publicly available.

Our methodology enables us to perform a thorough analysis of tweet geolocation, revealing insights into the best approaches for an accurate country-level location classifier for tweets.
We find that the use of a single feature like content, which is the most commonly used feature in previous work, does not suffice for an accurate classification of users by country and that the combination of multiple features leads to substantial improvement, outperforming the state-of-the-art real-time tweet geolocation classifier; this improvement is particularly manifest when using metadata like the user's self-reported location as well as the user's real name. We also perform a per-country analysis for the top 25 countries in terms of tweet volume, exploring how different features lead to optimal classification for different countries, as well as discussing limitations when dealing with some of the most challenging countries. We show that country-level classification of an unfiltered Twitter stream is challenging. It requires careful design of a classifier that uses an appropriate combination of features.
Our results at the country level are promising enough in the case of numerous countries, encouraging further research into finer-grained geolocation of global tweets. Cases where country-level geolocation is more challenging include English and Spanish speaking countries, which are harder to distinguish due to their numerous commonalities. Still, our experiments show that we can achieve F1 scores above 80\% in many of these cases given the choice of an appropriate combination of features, as well as an overall performance above 80\% in terms of both micro-accuracy and macro-accuracy for the top 25 countries.

\section{Related Work}

A growing body of research deals with the automated inference of demographic details of Twitter users \cite{mislove2011understanding}. Researchers have attempted to infer attributes of Twitter users such as age \cite{hughes2012tale,rao2010classifying}, gender \cite{burger2011discriminating,liu2013s,miller2012gender,rao2010classifying}, political orientation \cite{conover2011political,conover2011predicting,pennacchiotti2011democrats,pennacchiotti2011machine} or a range of social identities \cite{priante2016whoami}. Digging more deeply into the demographics of Twitter users, other researchers have attempted to infer socioeconomic demographics such as occupational class \cite{preoctiuc2015analysis}, income \cite{preoctiuc2015studying} and socioeconomic status \cite{lampos2016inferring}. Work by Huang et al. \cite{huang2014inferring} has also tried to infer the nationality of users; this work is different from that which we report here in that the country where the tweets were posted from, was already known.

What motivates the present study is the increasing interest in inferring the geographical location of either tweets or Twitter users \cite{ajao2015survey}. The automated inference of tweet location has been studied for different purposes, ranging from data journalism \cite{heravi2015tweet,middleton2016geoparsing} to public health \cite{dredze2013carmen}. As well as numerous different techniques, researchers have relied on different settings and pursued different objectives when conducting experiments. Table \ref{tab:related-work-summary} shows a summary of previous work reported in the scientific literature, outlining the features that each study used to classify tweets by location, the geographic scope of the study, the languages they dealt with, the classification granularity they tried to achieve and used for evaluation, and whether single tweets, aggregated multiple tweets and/or user history were used to train the classifier.

\begin{table*}
 \footnotesize
 \centering
 \begin{tabular}{| l | c | c | c | c | c |}
  \hline
  \textbf{Authors} & \textbf{Features} & \textbf{Geographic scope} & \textbf{Languages} & \textbf{Classif. granularity} & \textbf{Tweets/Users} \\
  \hline
  Eisenstein et al. \cite{eisenstein2010latent} & Tweet content & US only & All & Grid cells & Tweets \\
  \hline
  Cheng et al. \cite{cheng2010you} & Tweet content & US only & All & City-level & Users \\
  \hline
  Wing and Baldridge \cite{wing2011simple} & Tweet content & US only & All & Grid cells & Tweets \\
  \hline
  Roller et al. \cite{roller2012supervised} & Tweet content & US only & All & Grid cells & Tweets \\
  \hline
  Bo et al. \cite{bo2012geolocation} & Tweet content & Worldwide, 3.7k cities & English & City-level & Tweets \\
  \hline
  Chang et al. \cite{chang2012phillies} & Tweet content & US only & English & City-level & Users \\
  \hline
  Chen et al. \cite{chen2013interest} & Tweet content & Worldwide & English & City-level & Users \\
  \hline
  Jurgens \cite{jurgens2013s} & Social network & Worldwide & All & City-level & Users \\
  \hline
  Rodrigues et al. \cite{rodrigues2013uncovering} & Tweet content + social network & Brazil only, 3 cities & Portuguese & City-level & Users \\
  \hline
  Rout et al. \cite{rout2013s} & Social network & UK only & English & City-level & Users \\
  \hline
  Doran et al. \cite{doran2014accurate} & Tweet content & New York only & English & Grid cells & Tweets \\
  \hline
  Graham et al. \cite{graham2014world} & Tweet content & 4 metropolitan areas & 9 languages & City-level & Tweets \\
  \hline
  Han et al. \cite{han2014text} & Tweet content + 4 metadata & Worldwide, 3.1k cities & English & City and country & Users \\
  \hline
  Lee et al. \cite{lee2014twitter} & Tweet content & Manhattan only & English & Fine-grained location & Users \\
  \hline
  Mahmud et al. \cite{mahmud2014home} & Tweet content + user activity & US only & English & City-level & Users \\
  \hline
  Compton et al. \cite{compton2014geotagging} & Social network & Worldwide & All & City-level & Users \\
  \hline
  Krishnamurty et al. \cite{krishnamurthy2015knowledge} & Tweet content & US only & All & City-level & Users \\
  \hline
  Palpanas et al. \cite{palpanas2015fine} & Tweet content & Italy, 6 cities & English \& Italian & City-level & Tweets \\
  \hline
  Dredze et al. \cite{dredze2016geolocation} & Tweet content + 3 metadata & Worldwide, 3.7k cities & English & City and country & Tweets \\
  \hline
  \hline
  Present work & Tweet content + 7 metadata & Worldwide & All & Country-level & Tweets \\
  \hline
 \end{tabular}
 \caption{Characteristics of previous studies of automated geolocation of tweets or Twitter users. The present study, in the last row, represents the first attempt to deal with global tweets and in any language by using only features that are readily available within the body of a tweet or its metadata.}
 \label{tab:related-work-summary}
\end{table*}

Most of the previous studies on automated geolocation of tweets have assumed that the tweet stream includes only tweets from a specific country. The majority of these studies have focused on the United States, classifying tweets either at a city or state level. One of the earliest studies is that by Cheng et al. \cite{cheng2010you}, who introduced a probabilistic, content-based approach that identifies the most representative words of each of the major cities in the USA; these words are then used to classify new tweets. They incorporate different techniques to filter words, such as local and state-level filtering, classifying up to 51\% of Twitter users accurately within a 100 mile radius. Their approach, however, relies on making use of the complete history of a user, and was tested only for users with at least 1,000 tweets in their timeline.

Most of the other studies documented in the literature have also relied on tweet content, using different techniques such as topic modelling to find locally relevant keywords that reveal a user's likely location \cite{chang2012phillies,chen2013interest,cheng2010you,compton2014geotagging,han2014text,krishnamurthy2015knowledge,li2014fine,mahmud2014home,rodrigues2013uncovering}. Another widely used technique relies on the social network that a user is connected to, in order to infer a user's location from that of their followers and followees \cite{jurgens2013s,rodrigues2013uncovering,rout2013s}. While the approaches summarised will work well for certain applications, retrieving the tweet history for each user or the profile information of all of a user's followers and followees is not feasible in a real-time scenario. 
Hence, in this context, a classifier needs to deal with the additional challenge of having to rely only on the information that can be extracted from a single tweet.

Only a handful of studies have relied solely on the content of a single tweet to infer its location \cite{bo2012geolocation,doran2014accurate,eisenstein2010latent,graham2014world,palpanas2015fine,roller2012supervised,wing2011simple}. Again, most of these have actually worked on very restricted geographical areas, with tweets being limited to different regions, such as the United States \cite{eisenstein2010latent,wing2011simple}, four different cities \cite{graham2014world}, and New York only \cite{doran2014accurate}. Bo et al. \cite{bo2012geolocation} did focus on a broader geographical area, including 3.7k cities all over the world. Nevertheless, their study focused on a limited number of cities, disregarding other locations, and only classified tweets written in English.

When it comes to geolocation classification granularity, the majority of studies have aimed at city-level classification. While this provides fine-grained classification of tweets, it also means that a limited number of cities can be considered, ignoring other cities and towns. Only Han et al. \cite{han2014text} and Dredze et al. \cite{dredze2016geolocation} perform country-level classification, although they also restricted themselves to English language tweets posted from a limited number of cities. This means that tweets posted from cities other than the ones under consideration are removed from the stream, as are tweets written in other languages. In our study, we take as input the stream of tweets with content originating from any country and in any language, i.e. the entire tweet stream, to classify, at the country-level, each tweet according to its origin.

To date, the work by Han et al. \cite{han2014text} is the most relevant to our new study. They conducted a comprehensive study on how Twitter users can be geolocated by using different features of tweets. They analysed how location indicative words from a user's aggregated tweets can be used to geolocate the user. However, this requires collecting a user's history of tweets, which is not realistic in our real-time scenario. They also looked at how some metadata from tweets can be leveraged for classification, achieving slight improvements in performance, but again this is for a user's aggregated history. Finally, they looked at the temporality of tweets, using an old model to classify new tweets, finding that new tweets are more difficult to classify. This is an insightful study, which also motivates some of the settings and selection of classifiers in our own study; however, while an approach based on location indicative words may be very useful when looking at a user's aggregated tweets, it is rather limited when -- as in our case -- relying on a single tweet per user. Instead, our analysis of different tweet features for geolocating a tweet is based solely on its attributes as retrieved from the Twitter API. Dredze et al. \cite{dredze2016geolocation} followed an approach similar to ours when they looked at the utility of a model trained from past tweets, finding that the classification performance degrades for new tweets and that the trained model needs to be continually updated. Their study did not look into further details, such as whether some features are still useful for new tweets, however, and which our study analyses in more detail.

In summary, as far as we are aware, no previous work has dealt with the multiple features available within a tweet, as retrieved from the Twitter streaming API, to determine the location of a tweet posted from anywhere in the world.
We look at the suitability of eight tweet features for this purpose, both singly and combined, and experiment on two datasets collected within different time frames to measure the usefulness of an old model on new tweets.

\section{Datasets}

For training our classifier, we rely on the most widely adopted approach for the collection of a Twitter dataset with tweets categorised by location. This involves using the Twitter API endpoint that returns a stream of geolocated tweets posted from within one or more specified geographic bounding boxes\footnote{Twitter API's 'statuses/filter' endpoint: \url{https://dev.twitter.com/streaming/reference/post/statuses/filter}}. In our study, we set this bounding box to be the whole world (i.e., [-180,-90,180,90]) in order to retrieve tweets worldwide. This way, we collected streams of global geolocated tweets for two different week long periods: 4-11 October, 2014 (TC2014) and 22-28 October, 2015 (TC2015). This led to the collection of 31.7 million tweets in 2014 and 28.8 million tweets in 2015, which we adapt for our purposes as explained below.

Our raw datasets reflect the well-known fact that some Twitter users are far more prolific than others, which would introduce a bias in the evaluation if not dealt with. If our classifier has seen a user before, it is very likely that the user will tweet from the same country again. Hence, in order to ensure an unbiased evaluation of the tweet level classification, we de-duplicated users from our datasets, by randomly picking only one tweet from each user for TC2014. For TC2015, we also picked one tweet per user at random, but also removed users that were included in TC2014. This led to a collection of 4,155,763 geolocated tweets in TC2014 and 897,341 geolocated tweets in TC2015. 462,536 tweets were removed from the TC2015 dataset for belonging to users that also appeared in TC2014.

Having these tweets geolocated with the specific coordinates of the user's location, we then inferred the name of that location. For this, we used Nominatim\footnote{\url{http://wiki.openstreetmap.org/wiki/Nominatim}}, whose reverse geocoding feature enabled us to retrieve detailed information of the location pointed to by the coordinates given as input. From Nominatim's output, we made use of the country code in our experiments that aimed at country level classification of tweets. As a result, we had all the tweets in TC2014 and TC2015 categorised by country, which we then used as the ground truth for our classification experiments. It is worthwhile noting that the distribution of countries in TC2014 and TC2015 correlate highly with $r = 0.982$. This suggests that the distribution is stable and therefore we can focus our study on the usefulness of the model trained for different features for new tweets.

\begin{figure}[hbt]
 \centering
 \includegraphics[width=0.5\textwidth]{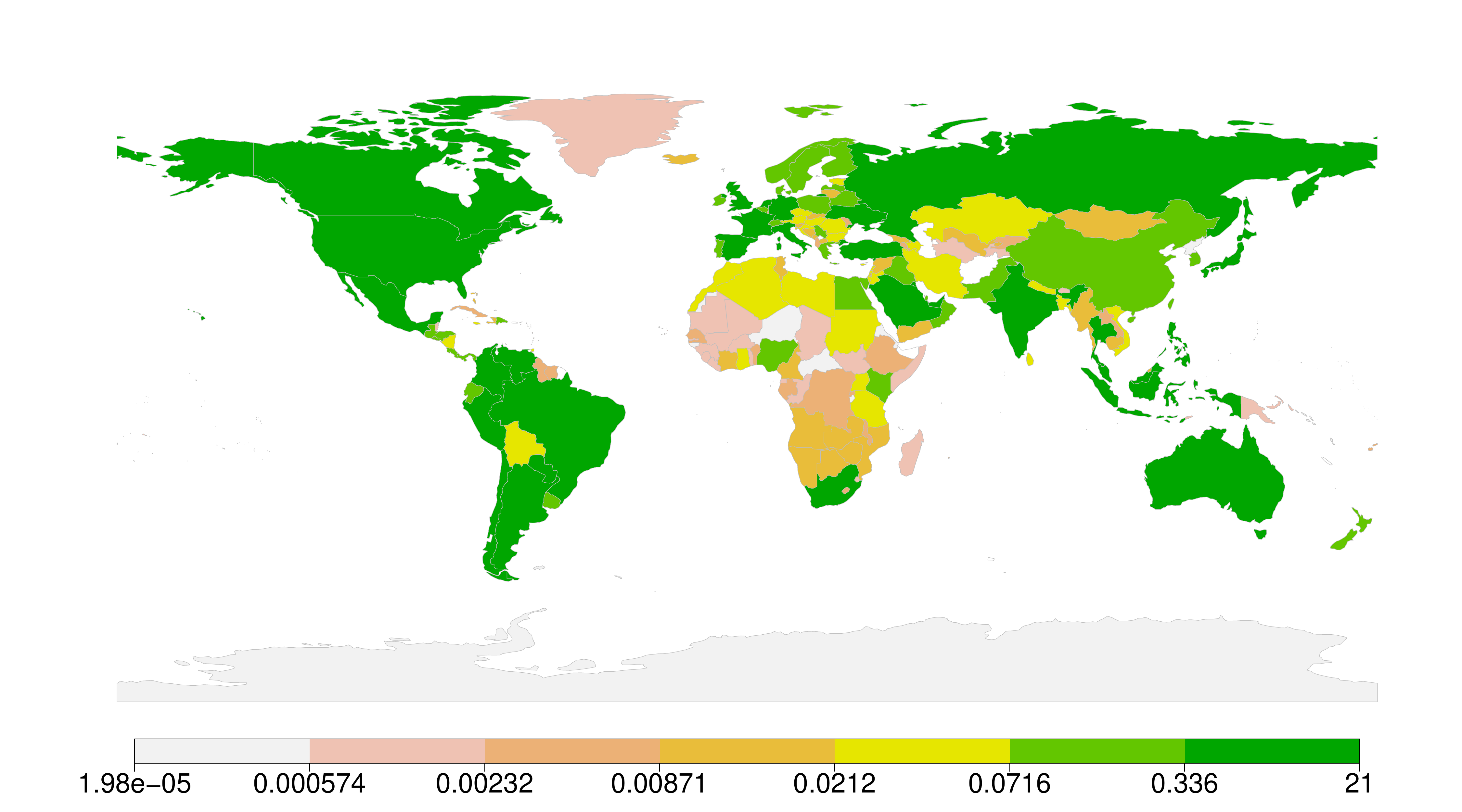}
 \caption{Prominence of countries in TC2014 and TC2015. Values in the legend represent percentages with respect to the entire dataset.}
 \label{fig:country-popularity}
\end{figure}

The more than 5 million tweets in these two datasets are categorised into 217 different countries. It is worthwhile mentioning that, as one would expect, the resulting datasets are clearly imbalanced, where only a few countries account for most of the tweets. The first country by number of tweets is the United States (20.99\%), followed by Indonesia (14.01\%) and Turkey (8.50\%). The 10 most prominent countries on Twitter in our datasets account for 72.98\% of the tweets, while the 25 most prominent countries account for 90.22\%. Figure \ref{fig:country-popularity} shows a heat map of popularity by country in our datasets.

The resulting datasets, both TC2014 and TC2015, are publicly available\footnote{Datasets, as well as details enabling reproducibility, are available through figshare: \url{https://figshare.com/articles/Tweet\_geolocation\_5m/3168529}}.

\section{Country-Level Location Classification for Tweets}
\label{sec:tcc}

In this study, we define the country-level location classification task as one in which, given a single tweet as input, a classifier has to determine the country of origin of the tweet. We argue for the sole use of the content and metadata provided in a single tweet\footnote{https://dev.twitter.com/overview/api/tweets}, which are accessible in a scenario where one wants to classify tweets by country in the tweet stream and in real-time. Most existing approaches have looked at the history of a Twitter user or the social network derivable from a user's followers and followees, which would not be feasible in our real-time scenario.

\subsection{Classification Techniques}
\label{ssec:classification-techniques}

We carried out the experimentation with a range of classifiers of different types: Support Vector Machines (SVM), Gaussian Naive Bayes, Multinomial Naive Bayes, Decision Trees, Random Forests and a Maximum Entropy classifier. They were tested in two different settings, one without balancing the weights of the different classes and the other by weighing the classes as the inverse of their frequency in the training set; the latter was tested as a means for dealing with the highly imbalanced data. The selection of these classifiers is in line with those used in the literature, especially with those tested by Han et al. \cite{han2014text}. This experimentation led to the selection of the weighed Maximum Entropy (MaxEnt) classifier as the most accurate. In the interest of space and focus, we only present results for this classifier.

Additionally, we compare our results with two baseline approaches. On the one hand, we used the Vowpal Wabbit classifier described by \cite{dredze2016geolocation}, a state-of-the-art real-time tweet geolocation classifier. On the other hand, we made use of the GeoNames geographical database\footnote{\url{http://www.geonames.org/}}, a commonly used approach in the literature. The user location, a string optionally specified by users in their profile settings, can be used here as input to the GeoNames database, which will return a likely location translated from that string. GeoNames provides a list of the most likely locations for a given string, based on either relevance or population, from which we took the first element. While GeoNames can be very effective for certain location names that are easy to map, the use of this feature is limited to users who opt to specify a non-empty location string in their settings (67.1\% in our datasets), and will fail with users whose location is not a valid country or city name (e.g., \textit{somewhere in the world}). The location specified in the user's profile has been used before to infer a user's location, although it is known to lead to low recall \cite{naaman2012study}. Here, we used this approach, using a database to translate user locations as a baseline, and explored whether, how, and to what extent a classifier can outperform it. For this baseline approach, we query GeoNames with the location string specified by the user and pick the first option output by the service. To make a fairer comparison with our classifiers, since GeoNames will not be able to determine the location for users with an empty location field, we default GeoNames' prediction for those tweets to be the majority country, i.e., the United States. This decision favours the baseline by assigning the most likely country and is also in line with the baseline approaches used in previous work \cite{han2014text}.

\subsection{Experiment Settings}

Within the TC2014 dataset, we created 10 different random distributions of the tweets for cross-validation, each having 50\% of the tweets for training, 25\% for development and 25\% for testing. The performance of the 10 runs on the test set were ultimately averaged to get the final performance value. The development set was used to determine the optimal parameters in each case, which are then used for the classification applied to the test set. In separate experiments, TC2015 was used as the test set, keeping the same subsets of TC2014 as training sets, to make the experiments comparable by using the same trained models and to assess the usefulness of year-old tweets to classify new tweets.

We created eight different classifiers, each of which used one of the following eight features available from a tweet as retrieved from a stream of the Twitter API:

\begin{enumerate}[leftmargin=*]
 \item \textit{User location (uloc):} This is the location the user specifies in their profile. While this feature might seem {\textit a priori} useful, it is somewhat limited as this is a free text field that users can leave empty, input a location name that is ambiguous or has typos, or a string that does not match with any specific locations (e.g., ``at home''). Looking at users' self-reported locations, Hecht et al. \cite{hecht2011tweets} found that 66\% report information that can be translated, accurately or inaccurately, to a geographic location, with the other 34\% being either empty or not geolocalisable.
 \item \textit{User language (ulang):} This is the user's self-declared user interface language. The interface language might be indicative of the user's country of origin; however, they might also have set up the interface in a different language, such as English, because it was the default language when they signed up or because the language of their choice is not available.
 \item \textit{Timezone (tz):} This indicates the time zone that the user has specified in their settings, e.g., ``Pacific Time (US \& Canada)''. When the user has specified an accurate time zone in their settings, it can be indicative of their country of origin; however, some users may have the default time zone in their settings, or they may use an equivalent time zone belonging to a different location (e.g., ``Europe/London'' for a user in Portugal). Also, Twitter's list of time zones does not include all countries.
 \item \textit{Tweet language (tlang):} The language in which a tweet is believed to be written is automatically detected by Twitter. It has been found to be accurate for major languages, but it leaves much to be desired for less widely used languages. Twitter's language identifier has also been found to struggle with multilingual tweets, where parts of a tweet are written in different languages \cite{zubiaga2015tweetlid}.
 \item \textit{Offset (offset):} This is the offset, with respect to UTC/GMT, that the user has specified in their settings. It is similar to the time zone, albeit more limited as it is shared with a number of countries.
 \item \textit{User name (name):} This is the name that the user specifies in their settings, which can be their real name, or an alternative name they choose to use. The name of a user can reveal, in some cases, their country of origin.
 \item \textit{User description (description):} This is a free text where a user can describe themselves, their interests, etc.
 \item \textit{Tweet content (content):} The text that forms the actual content of the tweet. The use of content has a number of caveats. One is that content might change over time, and therefore new tweets might discuss new topics that the classifiers have not seen before. Another caveat is that the content of the tweet might not be location-specific; in a previous study, Rakesh et al. \cite{rakesh2013location} found that the content of only 289 out of 10,000 tweets was location-specific.
\end{enumerate}

Figure \ref{fig:example-tweet} shows an example of a tweet and the eight features listed above. The features were treated in two different ways: the user location, name of the user, description and tweet content were represented using a bag of words approach, where each token represented a feature in the vector space model. The rest of the features, namely the user language, time zone, tweet language and offset, were represented by a single categorical value in the vector space model, given the limited number of values that the features can take. We used these eight features separately, as well as in different combinations with one another, in our experiments testing the ability to infer the country of origin of tweets. In separate experiments, we also append these features into single vectors to test different combinations of these features.

\begin{figure*}[tbh]
 \footnotesize
 \begin{framed}
  \noindent \{
  \begin{addmargin}[2em]{0pt}
   [text] $\rightarrow$ It is absolutely gorgeous outside. We will be delivering ice cream all day if you feel the need to not step out. \textbf{[content]}
  \end{addmargin}
  \begin{addmargin}[2em]{0pt}
   [lang] $\rightarrow$ en \textbf{[tlang]}
  \end{addmargin}
  \begin{addmargin}[2em]{0pt}
   [user] \{
  \end{addmargin}
  \begin{addmargin}[4em]{0pt}
   [utc\_offset] $\rightarrow$ -10800 \textbf{[offset]}
  \end{addmargin}
  \begin{addmargin}[4em]{0pt}
   [description] $\rightarrow$ \#FightForBigMike \textbf{[description]}
  \end{addmargin}
  \begin{addmargin}[4em]{0pt}
   [location] $\rightarrow$ FL \textbf{[location]}
  \end{addmargin}
  \begin{addmargin}[4em]{0pt}
   [lang] $\rightarrow$ en \textbf{[ulang]}
  \end{addmargin}
  \begin{addmargin}[4em]{0pt}
   [name] $\rightarrow$ John Smith \textbf{[name]}
  \end{addmargin}
  \begin{addmargin}[4em]{0pt}
   [time\_zone] $\rightarrow$ Atlantic Time (Canada) \textbf{[tz]}
  \end{addmargin}
  \begin{addmargin}[2em]{0pt}
   \}
  \end{addmargin}
  \noindent \}
 \end{framed}
 \caption{Example of a tweet and the 8 features that we used to infer the country of origin.}
 \label{fig:example-tweet}
\end{figure*}


\subsection{Evaluation}

We report three different performance values for each of the experiments: micro-accuracy, macro-accuracy and mean squared error (MSE). The accuracy values are computed as the result of dividing all the correctly classified instances by all the instances in the test set. The micro-accuracy is computed for the test set as a whole. For macro-accuracy, we compute the accuracy for each specific country in the test set, which are then averaged to compute the overall macro-accuracy. While the micro-accuracy measures the actual accuracy in the whole dataset, the macro-accuracy penalises the classifier that performs well only for the majority classes and rewards, instead,  classifiers that perform well across multiple categories. This is especially crucial in a case like ours where the categories are highly imbalanced.

The MSE is the average of the squared distance in kilometres between the predicted country and the actual, ground truth country, as shown in Equation \ref{eq:mse}.

\begin{equation}
 \frac{1}{n}\sum_{i=1}^n(\hat{Y_i} - Y_i)^2
 \label{eq:mse}
\end{equation}

In this computation, the distances between pairs of countries were calculated based on their centroids. We used the Countries of the World (COW) dataset produced by OpenGeonames.org to obtain the centroids of all countries. Having the latitude and longitude values of the centroids of all these countries, we then used the Haversine formula \cite{robusto1957cosine}, which accounts for the spheric shape when computing the distance between two points and is often used as an acceptable approximation to compute distances on the Earth. The Haversine distance between two points of a sphere each defined by its longitude and latitude is computed as shown in Equation \ref{eq:haversine}.

\begin{equation}
 \scriptsize
 d = 2 r \arcsin\left(\sqrt{\sin^2\left(\frac{\varphi_2 - \varphi_1}{2}\right) + \cos(\varphi_1) \cos(\varphi_2)\sin^2\left(\frac{\lambda_2 - \lambda_1}{2}\right)}\right)
 \label{eq:haversine}
\end{equation}

where $\varphi_1$ and $\varphi_2$ are the latitudes of point 1 and point 2, $\lambda_1$ and $\lambda_2$ are the longitudes of point 1 and point 2, and $r$ is the radius of the Earth, which is estimated to be 6,371 km.

\section{Classification Results}

In this section, we present results for different location classification experiments. First, we look at the performance of classifiers that use a single feature. Then, we present the results for classifiers combining multiple features. To conclude, we examine the results in more depth by looking at the performance by country, as well as error analysis.

\subsection{Single Feature}

Table \ref{tab:geonames} shows the results for the classification on the TC2014 dataset with two different approaches using GeoNames, one based on population (the most populous city is chosen when there are different options for a name) and one based on relevance (the city name that most resembles the input string). In this dataset, 65.82\% of the tweets have a non-empty string in the location field; for the rest of tweets, we pick the most popular country in the dataset as the output of the approach based on GeoNames. The table shows values of micro- and macro-accuracy.

There is no big difference between the two approaches based on GeoNames when we look at micro-accuracy. However, this accuracy is slightly better distributed across countries when we use the approach based on relevance, as can be seen from the macro-accuracy values. In what follows, we consider the relevance-based GeoNames approach as the baseline that solely relies on a database matching the user's profile location and compare with the use of classifiers that exploit additional features available in a tweet.

\begin{table}[hbt]
 \centering
 \begin{tabular}{| l | r | r | r |}
  \hline
  Feature & Microacc. & Macroacc. & MSE \\
  \hline
  population & \textbf{0.505} & 0.317 & 1505.661 \\
  \hline
  relevance & 0.504 & \textbf{0.342} & \textbf{1505.586} \\
  \hline
 \end{tabular}
 \caption{Classification results using GeoNames.}
 \label{tab:geonames}
\end{table}

\begin{figure*}[p]
 \centering
 \includegraphics[width=1.0\textwidth]{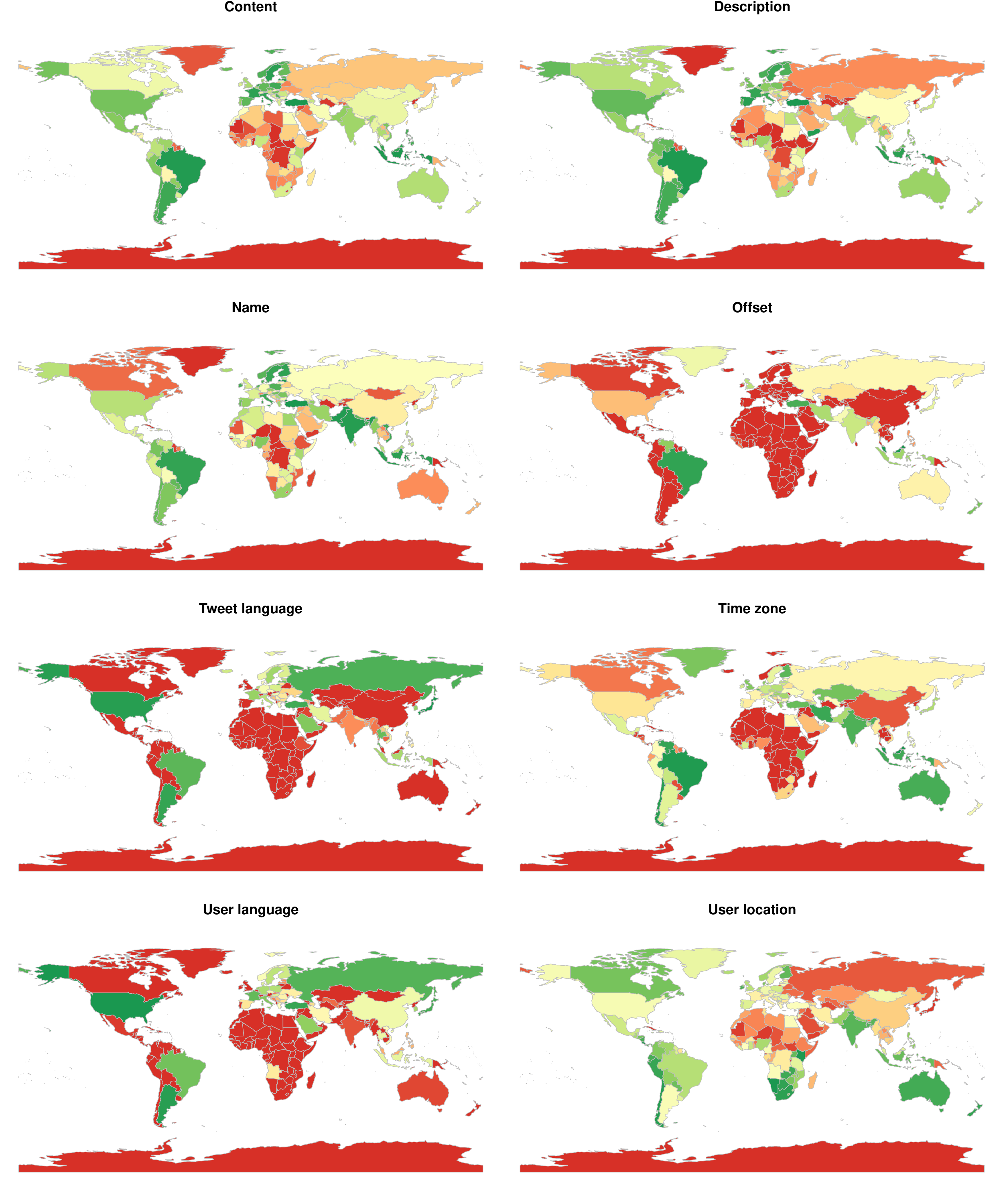}
 \caption{Accuracy by country for each of the eight features used alone in the classifier.}
 \label{fig:single-features-maxent-country-accuracies}
\end{figure*}

Table \ref{tab:single-maxent} shows the classification results, each case making use of only one of the eight features under study. This table includes performance values when we applied the classifier on both datasets, TC2014 and TC2015. The additional column, ``Diff.'', shows the relative difference in performance for each of these datasets, i.e., measuring the extent to which a model learned from the TC2014 dataset can still be applied to the TC2015 test set. Note that while higher values are desired for micro-accuracy and macro-accuracy, lower values are optimal for MSE.

\begin{table*}[htb]
 \centering
 \begin{tabular}{| l | r | r | r | r | r | r | r | r | r |}
  \hline
  \multicolumn{1}{| c |}{\multirow{2}{*}{\textbf{Feature}}} & \multicolumn{3}{ c |}{\textbf{Micro-accuracy}} & \multicolumn{3}{ c |}{\textbf{Macro-accuracy}} & \multicolumn{3}{ c |}{\textbf{MSE}} \\
  \cline{2-10}
  & \multicolumn{1}{ c |}{\textbf{TC2014}} & \multicolumn{1}{ c |}{\textbf{TC2015}} & \multicolumn{1}{ c |}{\textbf{Diff.}} & \multicolumn{1}{ c |}{\textbf{TC2014}} & \multicolumn{1}{ c |}{\textbf{TC2015}} & \multicolumn{1}{ c |}{\textbf{Diff.}} & \multicolumn{1}{ c |}{\textbf{TC2014}} & \multicolumn{1}{ c |}{\textbf{TC2015}} & \multicolumn{1}{ c |}{\textbf{Diff.}} \\
  \hline
  content & 0.503 & \textbf{0.588} & \textbf{+16.9\%} & 0.188 & 0.264 & \textbf{+40.4\%} & 1404.002 & \textbf{1148.264} & \textbf{-18.2\%} \\
  \hline
  description & 0.322 & 0.325 & +0.9\% & 0.096 & 0.095 & -1.0\% & 1870.311 & 1868.584 & -0.1\% \\
  \hline
  name & 0.232 & 0.232 & +0.0\% & 0.086 & 0.081 & -5.8\% & 2186.904 & 2190.848 & +0.2\% \\
  \hline
  offset & 0.267 & 0.233 & -12.7\% & 0.048 & 0.039 & -18.8\% & 2096.595 & 2173.044 & +3.6\% \\
  \hline
  tlang & \textbf{0.568} & 0.536 & -5.6\% & 0.107 & 0.088 & -17.8\% & \textbf{1156.279} & 1262.012 & +9.1\% \\
  \hline
  tz & 0.304 & 0.318 & +4.6\% & 0.123 & 0.118 & -4.1\% & 2013.270 & 1946.919 & -3.3\% \\
  \hline
  ulang & 0.547 & 0.525 & -4.0\% & 0.076 & 0.069 & -9.2\% & 1354.614 & 1468.346 & +8.4\% \\
  \hline
  uloc & 0.438 & 0.499 & +13.9\% & \textbf{0.374} & \textbf{0.370} & -1.1\% & 1669.383 & 1434.115 & -14.1\% \\
  \hline
 \end{tabular}
 \caption{Classification results with a Maximum Entropy classifier on a single feature for all the countries in TC2014 and TC2015. The last column, ``Diff.'', shows the relative difference in performance for each of these datasets.}
 \label{tab:single-maxent}
\end{table*}

\begin{table*}[htb]
 \centering
 \begin{tabular}{| l | r | r | r | r | r | r | r | r | r |}
  \hline
  \multicolumn{1}{| c |}{\multirow{2}{*}{\textbf{Feature}}} & \multicolumn{3}{ c |}{\textbf{Micro-accuracy}} & \multicolumn{3}{ c |}{\textbf{Macro-accuracy}} & \multicolumn{3}{ c |}{\textbf{MSE}} \\
  \cline{2-10}
  & \multicolumn{1}{ c |}{\textbf{TC2014}} & \multicolumn{1}{ c |}{\textbf{TC2015}} & \multicolumn{1}{ c |}{\textbf{Diff.}} & \multicolumn{1}{ c |}{\textbf{TC2014}} & \multicolumn{1}{ c |}{\textbf{TC2015}} & \multicolumn{1}{ c |}{\textbf{Diff.}} & \multicolumn{1}{ c |}{\textbf{TC2014}} & \multicolumn{1}{ c |}{\textbf{TC2015}} & \multicolumn{1}{ c |}{\textbf{Diff.}} \\
  \hline
  content & \textbf{0.632} & \textbf{0.667} & +5.5\% & \textbf{0.547} & \textbf{0.587} & +7.3\% & \textbf{926.810} & \textbf{838.722} & -9.5\% \\
  \hline
  description & 0.435 & 0.427 & 1.8\% & 0.390 & 0.385 & -1.3\% & 1318.187 & 1314.027 & -0.3\% \\
  \hline
  name & 0.452 & 0.446 & -1.3\% & 0.362 & 0.347 & -4.1\% & 1316.156 & 1305.311 & -0.8\% \\
  \hline
  offset & 0.340 & 0.318 & -6.5\% & 0.287 & 0.255 & -11.1\% & 1527.696 & 1543.421 & +1.0\% \\
  \hline
  tlang & 0.562 & 0.531 & -5.5\% & 0.404 & 0.357 & -11.6\% & 1120.293 & 1173.933 & +4.8\% \\
  \hline
  tz & 0.423 & 0.420 & -0.7\% & 0.389 & 0.395 & +1.5\% & 1320.393 & 1300.722 & -1.5\% \\
  \hline
  ulang & 0.542 & 0.520 & -4.1\% & 0.381 & 0.364 & -4.5\% & 1242.137 & 1283.080 & +3.3\% \\
  \hline
  uloc & 0.500 & 0.550 & \textbf{+10.0\%} & 0.507 & 0.552 & \textbf{+8.9\%} & 1184.321 & 1038.676 & \textbf{-12.3\%} \\
  \hline
 \end{tabular}
 \caption{Classification results with a Maximum Entropy classifier on a single feature for the top 25 countries in TC2014 and TC2015.}
 \label{tab:single-top25-maxent}
\end{table*}

If we look at the micro-accuracy scores, the results suggest that three approaches stand out over the rest. These are \textit{tweet content}, \textit{tweet language} and \textit{user language}, which are the only three approaches to get a micro-accuracy score above 0.5. However, these three approaches leave much to be desired when we evaluate them based on macro-accuracy scores, and therefore they fail to balance the classification well. Instead, the users' self-reported location (\textit{user location}) achieves the highest macro-accuracy scores, while micro-accuracy scores are only slightly lower. This is due to the fact that the classifier that only uses the user's profile location will be able to guess correctly a few cases for each country where users specify a correctly spelled, unambiguous location, but will fail to classify correctly the rest; hence the higher macro-accuracy is sensible according to these expectations. The MSE error rates suggest that \textit{tweet content} and \textit{tweet language} are the best in getting the most proximate classifications. We believe that this is due to the proximity of many countries that speak the same language (e.g., Germany and Austria, or Argentina and Chile), in which case the classifier that relies on tweet language or content will often choose a neighbouring country given the similarities they share in terms of topics and language. While most of these classifiers outperform the GeoNames baseline in terms of micro-accuracy, \textit{user location} is the only feature to beat the baseline in terms of macro-accuracy. However, the small improvement over the baseline suggests that alternative approaches are needed for a better balanced classification performance.

Figure \ref{fig:single-features-maxent-country-accuracies} shows a heat map with accuracy values of each of the features broken down by country. We observe the best distributed accuracy across countries is with the use of \textit{user location} as a feature. However, other features are doing significantly better classifying tweets that belong to some of the major countries such as the USA (better classified by \textit{tweet language} or \textit{user language}), Russia (better classified by \textit{tweet language}) or Brazil (better classified by \textit{tweet language}, \textit{user name} or \textit{tweet content}). This emphasises the necessity to explore further the differences between each country's characteristics.

As we noted above, a remarkable characteristic of our datasets (and the reality of Twitter itself) is the high imbalance in the distribution of tweets across countries, where a few countries account for a large majority of the tweets and many countries in the tail account for very few tweets. The fact that the classifier has to determine which of the 217 countries a tweet belongs to substantially complicates the task. To quantify this, and to explore the ability to boost performance on the countries with highest presence, we also performed classification experiments on the top 25 countries. These top 25 countries account for as many as 90.22\% of the tweets; consequently, being able to boost performance on these 25 countries, while assuming that the system will miss the rest, can make it a more achievable task where the overall performance gets improved.

To perform the classification on the top countries, we removed the tweets from countries that do not belong to the top 25 list from the training set. Including tweets from the remaining countries would add a noisy category to the training set, given the diversity of that new category. However, for obvious reasons, we cannot do the same for the test set. For the purposes of experimentation, we assign the rest of the tweets in the test set a different, 26th label, meaning that they belong to other countries. Our experiments on the top 25 countries will then have a training set with 25 categories to learn from and test sets with 26 categories, where the classifier will never predict the 26th category.

Table \ref{tab:single-top25-maxent} shows the results for the experiments on the top 25 countries. The overall tendency is very similar to that of the classifiers applied to all the countries in the world, with an expected overall boost in macro-accuracy values. However, we see a substantial improvement with the use of content as a feature, which now outperforms \textit{tweet language} in micro-accuracy scores as well as \textit{user location} in macro-accuracy scores. \textit{Tweet content} actually becomes the best performing feature with the reduced set of 25 countries. Classification on a reduced subset of countries can substantially boost performance, even assuming that part of the dataset will be misclassified. In fact, classification on this optimised setting outperforms by far the baseline using GeoNames. Not only does the top performing feature, \textit{tweet content}, improve its performance. Other features that performed poorly before, such as \textit{tweet language}, \textit{time zone} or \textit{user language}, perform significantly better, also outperforming the GeoNames baseline. This further motivates our subsequent goal of studying combinations of features to further boost the performance of the classifier applied to the top 25 countries.

\subsection{Feature Combinations}

\begin{table*}[tbh]
 \centering
 \begin{tabular}{| l | r | r | r | l | r | r | r |}
  \hline
  \multicolumn{8}{| c |}{\textbf{All countries}} \\
  \hline
  \multicolumn{4}{| c |}{\textbf{TC2014}} & \multicolumn{4}{ c |}{\textbf{TC2015}} \\
  \hline
  \multicolumn{1}{| c |}{\textbf{Feature}} & \multicolumn{1}{ c |}{\textbf{Micro.}} & \multicolumn{1}{ c |}{\textbf{Macro.}} & \multicolumn{1}{ c |}{\textbf{MSE}} & \multicolumn{1}{ c |}{\textbf{Feature}} & \multicolumn{1}{ c |}{\textbf{Micro.}} & \multicolumn{1}{ c |}{\textbf{Macro.}} & \multicolumn{1}{ c |}{\textbf{MSE}} \\
  \hline
  Dredze et al. \cite{dredze2016geolocation} & 0.666 & 0.122 & 862.792 & Dredze et al. \cite{dredze2016geolocation} & 0.636 & 0.116 & 956.997 \\
  \hline
  Best single feature & 0.568 & 0.374 & 1156.279 & Best single feature & 0.588 & 0.370 & 1148.264 \\
  \hline
  content-description-name- & \multirow{2}{*}{\textbf{0.889}} & \multirow{2}{*}{\textbf{0.452}} & \multirow{2}{*}{\textbf{244.106}} & content-description-name- & \multirow{2}{*}{\textbf{0.893}} & \multirow{2}{*}{\textbf{0.456}} & \multirow{2}{*}{\textbf{243.124}} \\
  tlang-tz-ulang-uloc &  &  &  & tlang-tz-ulang-uloc &  &  &  \\
  \hline
  Improvement & +56.5\% & +20.9\% & -78.9\% & Improvement & +51.9\% & +23.2\% & -78.9\% \\
  \hline

  \hline
  \multicolumn{8}{| c |}{\textbf{Top 25}} \\
  \hline
  \multicolumn{4}{| c |}{\textbf{TC2014}} & \multicolumn{4}{ c |}{\textbf{TC2015}} \\
  \hline
  \multicolumn{1}{| c |}{\textbf{Feature}} & \multicolumn{1}{ c |}{\textbf{Micro.}} & \multicolumn{1}{ c |}{\textbf{Macro.}} & \multicolumn{1}{ c |}{\textbf{MSE}} & \multicolumn{1}{ c |}{\textbf{Feature}} & \multicolumn{1}{ c |}{\textbf{Micro.}} & \multicolumn{1}{ c |}{\textbf{Macro.}} & \multicolumn{1}{ c |}{\textbf{MSE}} \\
  \hline
  Dredze et al. \cite{dredze2016geolocation} & 0.651 & 0.513 & 840.025 & Dredze et al. \cite{dredze2016geolocation} & 0.619 & 0.480 & 913.611 \\
  \hline
  Best single feature & 0.632 & 0.547 & 926.810 & Best single feature & 0.667 & 0.587 & 838.722 \\
  \hline
  content-description-name- & \multirow{2}{*}{\textbf{0.849}} & \multirow{2}{*}{\textbf{0.858}} & \multirow{2}{*}{\textbf{360.856}} & content-name-tlang- & \multirow{2}{*}{\textbf{0.837}} & \multirow{2}{*}{\textbf{0.853}} & \multirow{2}{*}{\textbf{385.807}} \\
  tlang-tz-ulang-uloc &  &  &  & tz-ulang-uloc &  &  &  \\
  \hline
  Improvement & +34.3\% & +56.9\% & -61.1\% & Improvement & +25.5\% & +45.3\% & -54.0\% \\
  \hline
 \end{tabular}
 \caption{Results for combinations of features, best performing single feature and the baseline classifier by Dredze et al. \cite{dredze2016geolocation}.}
 \label{tab:combined-maxent}
\end{table*}

Having seen that different features give rise to gains in different ways, testing the performance of combinations of multiple features seemed like a wise option. We performed these combinations of features by appending the vectors for each of the features into a single vector. We tested all 255 possible combinations using the eight features under study. We only report the best performing combinations here in the interest of space and clarity.

Table \ref{tab:combined-maxent} shows the best combination in each case for the TC2014 and TC2015 datasets, as well as for the classifiers that consider all the countries in the datasets and only the top 25 countries. The table also shows the performance of the best single feature as well as the baseline classifier by \cite{dredze2016geolocation} to facilitate comparison, as well as the improvement in performance when using a combination of features over that of a single feature. We observe that the selection of an appropriate combination of features can actually lead to a substantial increase in terms of all micro-accuracy, macro-accuracy and MSE. These improvements are especially remarkable when we look at the MSE scores, where the improvement is always above 50\%. Improvements in terms of micro-accuracy and macro-accuracy scores are also always above 20\%, but are especially high for micro-accuracy (50\%+) when we classify for all the countries, and for macro-accuracy (40\%+) when we classify for the top 25 countries. These results suggest that the use of a single feature, as it is the case with most previous work using e.g. only \textit{tweet content}, can be substantially improved by using more features. In fact, our results suggest that the combination of many features is usually best; we need to combine seven of the eight features (all but offset) in three of the cases, and six features in the other case (all but description and offset). As a result, we get performance values above 85\% in terms of macro-accuracy for the top 25 countries. These performance scores are also remarkably higher than those of the classifier by \cite{dredze2016geolocation}, both in terms of micro- and macro-accuracy.

Interestingly, the combination of features has led to a significant improvement in performance, with a better balance across countries. To complement this analysis, we believe it is important to understand the differences among countries. Will different sets of features be useful for an accurate classification for each country? Are we perhaps doing very well for some countries with certain combinations, but that combination, is in turn, bad for other countries? To explore this further, we now take a closer look at the performance broken down by country.

\subsection{Breakdown of Countries}

Given the remarkable differences among countries we observed (Figure \ref{fig:single-features-maxent-country-accuracies}) when exploring how different features are useful for different countries, we take a closer look at the performance of different classifiers for each of the top 25 countries. As we are now looking at each country separately, we use precision, recall and F1 scores as more appropriate evaluation measures that better capture the extent to which a country's tweets are being correctly categorised. We look at the best combination of features for each country in terms of F1 score and analyse the set of features that lead to the best performance in each case. We show the results of this analysis in Table \ref{tab:accuracy-by-country}.

  \begin{table*}[htbp]
   \scriptsize
   \centering
   \begin{tabular}{| l | p{0.1cm} | p{0.1cm} | p{0.1cm} | p{0.1cm} | p{0.1cm} | p{0.1cm} | p{0.1cm} | p{0.1cm} | r | r | r | p{0.1cm} | p{0.1cm} | p{0.1cm} | p{0.1cm} | p{0.1cm} | p{0.1cm} | p{0.1cm} | p{0.1cm} | r | r | r | r |}
    \hline
    & \multicolumn{11}{ c |}{\textbf{TC2014}} & \multicolumn{11}{ c |}{\textbf{TC2015}} &  \\
    \cline{1-23}
    \multicolumn{1}{| c |}{\textbf{Country}} & \multicolumn{8}{ c |}{\textbf{Best SVM combination}} & \multicolumn{3}{ c |}{\textbf{Performance}} & \multicolumn{8}{ c |}{\textbf{Best SVM combination}} & \multicolumn{3}{ c |}{\textbf{Performance}} &  \\
    \cline{1-23}
    & \rotatebox{90}{content} & \rotatebox{90}{description} & \rotatebox{90}{name} & \rotatebox{90}{offset} & \rotatebox{90}{tlang} & \rotatebox{90}{tz} & \rotatebox{90}{ulang} & \rotatebox{90}{uloc} & \multicolumn{1}{ c |}{\textbf{P}} & \multicolumn{1}{ c |}{\textbf{R}} & \multicolumn{1}{ c |}{\textbf{F1}} & \rotatebox{90}{content} & \rotatebox{90}{description} & \rotatebox{90}{name} & \rotatebox{90}{offset} & \rotatebox{90}{tlang} & \rotatebox{90}{tz} & \rotatebox{90}{ulang} & \rotatebox{90}{uloc} & \multicolumn{1}{ c |}{\textbf{P}} & \multicolumn{1}{ c |}{\textbf{R}} & \multicolumn{1}{ c |}{\textbf{F1}} & \multicolumn{1}{ c |}{\textbf{Diff.}} \\
    \hline
    Turkey & \cellcolor{black!50} &  & \cellcolor{black!50} &  & \cellcolor{black!50} & \cellcolor{black!50} & \cellcolor{black!50} & \cellcolor{black!50} & 0.973 & 0.988 & 0.980 & \cellcolor{green!100} &  & \cellcolor{green!100} &  & \cellcolor{green!100} &  & \cellcolor{green!100} & \cellcolor{green!100} & 0.973 & 0.990 & 0.982 & +0.2\% \\
    \hline
    Indonesia & \cellcolor{black!50} & \cellcolor{black!50} & \cellcolor{black!50} & \cellcolor{black!50} & \cellcolor{black!50} & \cellcolor{black!50} & \cellcolor{black!50} & \cellcolor{black!50} & 0.974 & 0.976 & 0.975 & \cellcolor{green!100} &  & \cellcolor{green!100} &  &  & \cellcolor{green!100} & \cellcolor{green!100} & \cellcolor{green!100} & 0.974 & 0.976 & 0.975 & +0.0\% \\
    \hline
    Brazil & \cellcolor{black!50} & \cellcolor{black!50} & \cellcolor{black!50} & \cellcolor{black!50} & \cellcolor{black!50} & \cellcolor{black!50} & \cellcolor{black!50} & \cellcolor{black!50} & 0.948 & 0.989 & 0.968 & \cellcolor{green!100} &  & \cellcolor{green!100} &  & \cellcolor{green!100} & \cellcolor{green!100} & \cellcolor{green!100} & \cellcolor{green!100} & 0.919 & 0.982 & 0.949 & -2.0\% \\
    \hline
    Japan &  &  &  &  & \cellcolor{black!50} & \cellcolor{black!50} & \cellcolor{black!50} &  & 0.955 & 0.977 & 0.965 & \cellcolor{red!100} &  &  &  & \cellcolor{red!100} &  & \cellcolor{red!100} &  & 0.949 & 0.969 & 0.959 & -0.6\% \\
    \hline
    Thailand & \cellcolor{black!50} & \cellcolor{black!50} & \cellcolor{black!50} & \cellcolor{black!50} & \cellcolor{black!50} & \cellcolor{black!50} & \cellcolor{black!50} & \cellcolor{black!50} & 0.924 & 0.949 & 0.936 & \cellcolor{green!100} &  & \cellcolor{green!100} &  & \cellcolor{green!100} & \cellcolor{green!100} & \cellcolor{green!100} & \cellcolor{green!100} & 0.914 & 0.932 & 0.923 & -1.4\% \\
    \hline
    USA & \cellcolor{black!50} & \cellcolor{black!50} & \cellcolor{black!50} & \cellcolor{black!50} & \cellcolor{black!50} & \cellcolor{black!50} & \cellcolor{black!50} & \cellcolor{black!50} & 0.889 & 0.945 & 0.916 & \cellcolor{green!100} &  & \cellcolor{green!100} & \cellcolor{green!100} &  & \cellcolor{green!100} & \cellcolor{green!100} & \cellcolor{green!100} & 0.864 & 0.933 & 0.897 & -2.1\% \\
    \hline
    Malaysia & \cellcolor{black!50} & \cellcolor{black!50} & \cellcolor{black!50} & \cellcolor{black!50} &  & \cellcolor{black!50} &  & \cellcolor{black!50} & 0.840 & 0.909 & 0.873 & \cellcolor{green!100} &  & \cellcolor{green!100} &  &  & \cellcolor{green!100} &  & \cellcolor{green!100} & 0.876 & 0.930 & 0.902 & +3.3\% \\
    \hline
    Italy & \cellcolor{black!50} & \cellcolor{black!50} & \cellcolor{black!50} &  &  &  & \cellcolor{black!50} & \cellcolor{black!50} & 0.847 & 0.894 & 0.870 & \cellcolor{green!100} &  & \cellcolor{green!100} &  &  &  & \cellcolor{green!100} & \cellcolor{green!100} & 0.828 & 0.873 & 0.850 & -2.3\% \\
    \hline
    Argentina & \cellcolor{black!50} & \cellcolor{black!50} & \cellcolor{black!50} & \cellcolor{black!50} & \cellcolor{black!50} & \cellcolor{black!50} & \cellcolor{black!50} & \cellcolor{black!50} & 0.804 & 0.938 & 0.865 & \cellcolor{green!100} &  & \cellcolor{green!100} &  &  & \cellcolor{green!100} &  & \cellcolor{green!100} & 0.828 & 0.902 & 0.863 & -0.2\% \\
    \hline
    Spain & \cellcolor{black!50} & \cellcolor{black!50} & \cellcolor{black!50} & \cellcolor{black!50} &  & \cellcolor{black!50} & \cellcolor{black!50} & \cellcolor{black!50} & 0.815 & 0.917 & 0.863 & \cellcolor{green!100} &  & \cellcolor{green!100} &  &  & \cellcolor{green!100} & \cellcolor{green!100} & \cellcolor{green!100} & 0.728 & 0.897 & 0.804 & -6.8\% \\
    \hline
    France & \cellcolor{black!50} &  &  & \cellcolor{black!50} & \cellcolor{black!50} &  & \cellcolor{black!50} & \cellcolor{black!50} & 0.797 & 0.929 & 0.858 & \cellcolor{blue!100} &  &  & \cellcolor{blue!100} & \cellcolor{blue!100} &  & \cellcolor{blue!100} & \cellcolor{blue!100} & 0.706 & 0.862 & 0.776 & -9.6\% \\
    \hline
    Philippines & \cellcolor{black!50} & \cellcolor{black!50} & \cellcolor{black!50} & \cellcolor{black!50} &  &  & \cellcolor{black!50} & \cellcolor{black!50} & 0.793 & 0.884 & 0.836 & \cellcolor{red!100} &  & \cellcolor{red!100} & \cellcolor{red!100} & \cellcolor{red!100} &  & \cellcolor{red!100} & \cellcolor{red!100} & 0.848 & 0.880 & 0.864 & +3.3\% \\
    \hline
    Russia &  &  &  & \cellcolor{black!50} & \cellcolor{black!50} & \cellcolor{black!50} & \cellcolor{black!50} & \cellcolor{black!50} & 0.714 & 0.985 & 0.828 & \cellcolor{red!100} &  &  & \cellcolor{red!100} & \cellcolor{red!100} & \cellcolor{red!100} & \cellcolor{red!100} & \cellcolor{red!100} & 0.674 & 0.966 & 0.794 & -4.1\% \\
    \hline
    UK & \cellcolor{black!50} & \cellcolor{black!50} & \cellcolor{black!50} &  &  & \cellcolor{black!50} &  & \cellcolor{black!50} & 0.751 & 0.879 & 0.810 & \cellcolor{red!100} &  & \cellcolor{red!100} &  &  & \cellcolor{red!100} &  & \cellcolor{red!100} & 0.673 & 0.857 & 0.754 & -6.9\% \\
    \hline
    Chile & \cellcolor{black!50} & \cellcolor{black!50} & \cellcolor{black!50} & \cellcolor{black!50} & \cellcolor{black!50} &  & \cellcolor{black!50} & \cellcolor{black!50} & 0.788 & 0.830 & 0.809 & \cellcolor{green!100} &  & \cellcolor{green!100} & \cellcolor{green!100} &  &  & \cellcolor{green!100} & \cellcolor{green!100} & 0.735 & 0.833 & 0.781 & -3.5\% \\
    \hline
    Mexico & \cellcolor{black!50} & \cellcolor{black!50} & \cellcolor{black!50} &  & \cellcolor{black!50} & \cellcolor{black!50} & \cellcolor{black!50} & \cellcolor{black!50} & 0.736 & 0.864 & 0.795 & \cellcolor{green!100} &  & \cellcolor{green!100} &  &  & \cellcolor{green!100} &  & \cellcolor{green!100} & 0.778 & 0.874 & 0.823 & +3.5\% \\
    \hline
    Netherlands & \cellcolor{black!50} & \cellcolor{black!50} & \cellcolor{black!50} &  &  &  & \cellcolor{black!50} & \cellcolor{black!50} & 0.721 & 0.880 & 0.793 & \cellcolor{green!100} &  & \cellcolor{green!100} &  &  &  &  & \cellcolor{green!100} & 0.568 & 0.787 & 0.660 & -16.8\% \\
    \hline
    Venezuela & \cellcolor{black!50} & \cellcolor{black!50} & \cellcolor{black!50} & \cellcolor{black!50} &  &  & \cellcolor{black!50} & \cellcolor{black!50} & 0.723 & 0.831 & 0.773 & \cellcolor{green!100} &  & \cellcolor{green!100} & \cellcolor{green!100} &  &  & \cellcolor{green!100} & \cellcolor{green!100} & 0.755 & 0.841 & 0.795 & +2.8\% \\
    \hline
    Colombia & \cellcolor{black!50} & \cellcolor{black!50} & \cellcolor{black!50} &  & \cellcolor{black!50} & \cellcolor{black!50} & \cellcolor{black!50} & \cellcolor{black!50} & 0.686 & 0.859 & 0.763 & \cellcolor{green!100} &  & \cellcolor{green!100} &  &  & \cellcolor{green!100} &  & \cellcolor{green!100} & 0.677 & 0.851 & 0.754 & -1.2\% \\
    \hline
    India & \cellcolor{black!50} & \cellcolor{black!50} & \cellcolor{black!50} &  &  & \cellcolor{black!50} &  & \cellcolor{black!50} & 0.614 & 0.859 & 0.716 & \cellcolor{green!100} &  & \cellcolor{green!100} &  &  & \cellcolor{green!100} &  & \cellcolor{green!100} & 0.681 & 0.846 & 0.755 & +5.4\% \\
    \hline
    Saudi Arabia &  &  &  &  &  &  & \cellcolor{black!50} & \cellcolor{black!50} & 0.636 & 0.793 & 0.705 &  &  &  &  &  &  & \cellcolor{blue!100} & \cellcolor{blue!100} & 0.445 & 0.745 & 0.557 & -21.0\% \\
    \hline
    Australia & \cellcolor{black!50} & \cellcolor{black!50} & \cellcolor{black!50} & \cellcolor{black!50} &  &  &  & \cellcolor{black!50} & 0.651 & 0.753 & 0.698 & \cellcolor{red!100} &  & \cellcolor{red!100} & \cellcolor{red!100} & \cellcolor{red!100} &  &  & \cellcolor{red!100} & 0.651 & 0.799 & 0.717 & +2.7\% \\
    \hline
    Canada &  &  &  & \cellcolor{black!50} & \cellcolor{black!50} & \cellcolor{black!50} &  & \cellcolor{black!50} & 0.775 & 0.586 & 0.667 & \cellcolor{red!100} &  &  & \cellcolor{red!100} & \cellcolor{red!100} &  & \cellcolor{red!100} & \cellcolor{red!100} & 0.691 & 0.744 & 0.717 & +7.5\% \\
    \hline
    South Africa & \cellcolor{black!50} & \cellcolor{black!50} & \cellcolor{black!50} &  &  & \cellcolor{black!50} &  & \cellcolor{black!50} & 0.568 & 0.801 & 0.665 & \cellcolor{red!100} &  & \cellcolor{red!100} & \cellcolor{red!100} &  & \cellcolor{red!100} &  & \cellcolor{red!100} & 0.682 & 0.807 & 0.739 & +11.1\% \\
    \hline
    Germany & \cellcolor{black!50} & \cellcolor{black!50} & \cellcolor{black!50} &  &  &  & \cellcolor{black!50} & \cellcolor{black!50} & 0.454 & 0.731 & 0.560 & \cellcolor{red!100} & \cellcolor{red!100} & \cellcolor{red!100} & \cellcolor{red!100} &  &  & \cellcolor{red!100} &  & 0.497 & 0.709 & 0.584 & +4.3\% \\
    \hline
    \hline
    TOTAL & 21 & 19 & 20 & 14 & 13 & 16 & 19 & 24 & -- & -- & -- & 23 & 1 & 20 & 10 & 9 & 13 & 16 & 24 & -- & -- & -- & -- \\
    \hline
   \end{tabular}
   \caption{Results broken down by country for the top 25 countries. The color code represents how the best sets of features for TC2015 compare to those for TC2014 (blue: countries where the same set of features works best for TC2014 and TC2015; green: countries where a reduced set of features from TC2014 works best for TC2015; red: countries where new features, not used in the best approach for TC2014, works best for TC2015).}
   \label{tab:accuracy-by-country}
  \end{table*}

The results show that very different approaches lead to optimal results for each country, revealing the different features that characterise each country. One striking observation we make from the ranking of country accuracies is that seven of the top eight ranking countries have unique characteristics, especially when it comes to language; except for the USA, these countries have a language that is not shared with any other country in the list. Interestingly, the best approach for most of these countries include either or both of \textit{tweet language} or \textit{user language}. When it comes to \textit{user language}, this means that users in these countries have a strong inclination towards setting the user interface in their own language instead of the default language. In the case of \textit{tweet language}, this mainly reflects a combination of two things, one being that users in these countries tend to tweet mostly in their own language, while the other is that Twitter's language identifier is very accurate in these cases. Further down in the list, we see the Spanish and English speaking countries, which seem to be harder to classify because of the numerous commonalities with one another, both in terms of language as well as in terms of content, given their cultural and geographical proximity.

All of the top 25 countries actually benefit from a combination of features, as there is no single case in which the use of only one feature performs best. Most of the countries in fact benefit from combining four or more features, with the only exceptions being Saudi Arabia --two features-- and Japan --three features. Looking at the utility of features (see last row of the table showing totals), the features that are useful for TC2014 in most of the cases include \textit{user location}, \textit{tweet content} and \textit{user name}, while  \textit{offset} and \textit{tweet language} are the least useful. When we look at the combinations that perform best for new tweets --i.e. TC2015--, we see that in the majority of the cases the optimal combination is a reduced subset of that for TC2014 (green rows). This suggests that there are some features that perform well when classifying tweets from the same time frame as the training data, but whose performance drops when applied to new collections of tweets. However, one can get comparable performance when the right combination of features is chosen. As our results suggest, the features whose utility tends to fade include especially \textit{user description}, with a remarkable drop from 19 to 1 case where it is useful, but also to a lesser extent \textit{tweet language}, \textit{offset}, \textit{time zone} and \textit{user language}. On the other hand, \textit{tweet content}, \textit{user name} and \textit{user location} are the features that are as useful when applied to new tweets.

Finally, looking at the performance difference of countries in TC2014 and that in TC2015, there is no big gap in most of the cases and the differences are mostly within $\pm$5\%. However, there are a few cases where the performance drops drastically when we apply the classifier on the new dataset. This is the case of Saudi Arabia, Netherlands and France, whose performance in TC2015 drops between 9\% and 21\% from that in TC2014. The highest improvement occurs for Germany, India and South Africa, with increases in performance in TC2014 that range between 4\% and 11\%.

\subsection{Error Analysis}

To shed some light on the reasons why some countries are not classified as accurately, we looked at the errors that the classifiers are making. Overall, if we put together all correct classifications by any of the classifiers, we would be able to get a micro-accuracy of up to 99.1\% as an upper bound estimation for the tweets that belong to one of the top 25 countries. This raises expectations in that nearly all users can be accurately classified in some way by using the right classifier. However, many countries share similar (or common) characteristics, which often leads to mistakes between those countries. To better understand this, we look at the confusion matrix for the top 25 countries.

The confusion matrix in Table \ref{tab:confusion-matrix} shows the aggregated misclassifications for all the 255 classifiers applied to the top 25 countries. The values highlighted in grey refer to correct guesses (diagonal). In red, we highlight misclassifications exceeding 10\% of a country's tweets, in orange those exceeding 5\% and in yellow those exceeding 2\%.

  \begin{sidewaystable*}[p]
   \footnotesize
   \centering
   \begin{tabular}{| l | r | r | r | r | r | r | r | r | r | r | r | r | r | r | r | r | r | r | r | r | r | r | r | r | r |}
    \hline
     & ar & au & br & ca & cl & co & de & es & fr & gb & id & in & it & jp & mx & my & nl & ph & ru & sa & th & tr & us & ve & za \\
    \hline
    ar & \cellcolor{gray!50}\textbf{.762} & .006 & \cellcolor{yellow!75}.022 & .003 & .019 & .018 & .007 & \cellcolor{orange!75}.065 & .004 & .004 & .003 & .005 & .005 & .006 & \cellcolor{yellow!75}.022 & .003 & .002 & .003 & .003 & .012 & .002 & .001 & .010 & .013 & .001 \\
    \hline
    au & .002 & \cellcolor{gray!50}\textbf{.603} & .005 & .017 & .001 & .002 & .008 & .006 & .006 & \cellcolor{orange!75}.078 & .015 & .017 & .006 & .010 & .002 & \cellcolor{yellow!75}.020 & .004 & \cellcolor{yellow!75}.022 & .007 & .014 & .007 & .003 & \cellcolor{red!75}.138 & .001 & .010 \\
    \hline
    br & .007 & .005 & \cellcolor{gray!50}\textbf{.898} & .002 & .003 & .001 & .006 & .007 & .004 & .004 & .003 & .004 & .004 & .007 & .002 & .003 & .002 & .003 & .003 & .011 & .002 & .001 & .012 & .001 & .003 \\
    \hline
    ca & .002 & .017 & .008 & \cellcolor{gray!50}\textbf{.434} & .001 & .003 & .007 & .005 & \cellcolor{yellow!75}.023 & \cellcolor{yellow!75}.037 & .011 & .015 & .005 & .008 & .007 & .008 & .003 & .015 & .006 & .014 & .006 & .003 & \cellcolor{red!75}.354 & .002 & .009 \\
    \hline
    cl & \cellcolor{red!75}.104 & .005 & .018 & .003 & \cellcolor{gray!50}\textbf{.659} & \cellcolor{yellow!75}.027 & .008 & \cellcolor{orange!75}.054 & .004 & .004 & .002 & .004 & .005 & .003 & \cellcolor{yellow!75}.034 & .003 & .001 & .003 & .003 & .010 & .001 & .001 & .018 & \cellcolor{yellow!75}.021 & .001 \\
    \hline
    co & \cellcolor{orange!75}.063 & .005 & .004 & .003 & .012 & \cellcolor{gray!50}\textbf{.657} & .006 & \cellcolor{orange!75}.060 & .004 & .004 & .002 & .005 & .004 & .005 & \cellcolor{orange!75}.082 & .003 & .002 & .003 & .003 & .013 & .001 & .001 & \cellcolor{yellow!75}.031 & \cellcolor{yellow!75}.029 & .001 \\
    \hline
    de & .005 & .011 & .011 & .007 & .002 & .003 & \cellcolor{gray!50}\textbf{.624} & \cellcolor{yellow!75}.024 & .016 & \cellcolor{yellow!75}.048 & .013 & .011 & .016 & .010 & .006 & .009 & .019 & .008 & .018 & \cellcolor{yellow!75}.025 & .006 & \cellcolor{yellow!75}.034 & \cellcolor{orange!75}.062 & .002 & .012 \\
    \hline
    es & \cellcolor{orange!75}.056 & .004 & .006 & .003 & .007 & .015 & .008 & \cellcolor{gray!50}\textbf{.758} & .008 & .018 & .003 & .004 & .007 & .005 & \cellcolor{yellow!75}.022 & .003 & .007 & .003 & .004 & .013 & .002 & .005 & .017 & .016 & .004 \\
    \hline
    fr & .004 & .007 & .007 & .007 & .001 & .002 & .009 & .016 & \cellcolor{gray!50}\textbf{.783} & \cellcolor{yellow!75}.026 & .006 & .006 & .008 & .011 & .004 & .006 & .010 & .006 & .007 & .018 & .004 & .012 & \cellcolor{yellow!75}.032 & .002 & .007 \\
    \hline
    gb & .002 & .017 & .003 & .012 & .001 & .001 & .008 & .010 & .007 & \cellcolor{gray!50}\textbf{.705} & .006 & .014 & .006 & .005 & .002 & .009 & .006 & .008 & .005 & .017 & .004 & .003 & \cellcolor{red!75}.136 & .001 & .012 \\
    \hline
    id & .001 & .005 & .001 & .002 & .000 & .000 & .004 & .003 & .001 & .004 & \cellcolor{gray!50}\textbf{.876} & .007 & .002 & .005 & .001 & .016 & .002 & .010 & .004 & .010 & .011 & .001 & \cellcolor{yellow!75}.030 & .001 & .003 \\
    \hline
    in & .001 & .016 & .002 & .010 & .000 & .001 & .008 & .005 & .004 & \cellcolor{yellow!75}.026 & .017 & \cellcolor{gray!50}\textbf{.696} & .004 & .009 & .001 & .014 & .003 & .011 & .009 & .032 & .008 & .003 & \cellcolor{red!75}.105 & .001 & .014 \\
    \hline
    it & .007 & .008 & .009 & .005 & .001 & .002 & .011 & \cellcolor{yellow!75}.022 & .012 & \cellcolor{yellow!75}.027 & .007 & .007 & \cellcolor{gray!50}\textbf{.754} & .008 & .004 & .004 & .009 & .008 & .009 & \cellcolor{yellow!75}.022 & .003 & .011 & \cellcolor{yellow!75}.039 & .003 & .007 \\
    \hline
    jp & .001 & .010 & .003 & .003 & .000 & .000 & .006 & .005 & .004 & .004 & .006 & .009 & .004 & \cellcolor{gray!50}\textbf{.830} & .001 & .013 & .001 & .012 & \cellcolor{yellow!75}.020 & \cellcolor{yellow!75}.042 & .010 & .002 & .011 & .002 & .001 \\
    \hline
    mx & \cellcolor{orange!75}.054 & .005 & .004 & .005 & .010 & \cellcolor{yellow!75}.047 & .007 & \cellcolor{orange!75}.055 & .004 & .005 & .004 & .006 & .004 & .005 & \cellcolor{gray!50}\textbf{.682} & .004 & .002 & .005 & .004 & .012 & .003 & .001 & \cellcolor{orange!75}.051 & .019 & .002 \\
    \hline
    my & .000 & .008 & .001 & .003 & .000 & .000 & .004 & .002 & .002 & .009 & \cellcolor{orange!75}.062 & .011 & .001 & .008 & .001 & \cellcolor{gray!50}\textbf{.768} & .002 & \cellcolor{yellow!75}.038 & .006 & .011 & .008 & .002 & \cellcolor{yellow!75}.049 & .001 & .005 \\
    \hline
    nl & .003 & .007 & .007 & .005 & .001 & .002 & .015 & .014 & .011 & \cellcolor{orange!75}.052 & .011 & .008 & .008 & .006 & .003 & .005 & \cellcolor{gray!50}\textbf{.735} & .006 & .006 & .012 & .004 & .012 & \cellcolor{yellow!75}.049 & .004 & .015 \\
    \hline
    ph & .001 & .015 & .002 & .008 & .000 & .001 & .006 & .005 & .004 & .015 & \cellcolor{yellow!75}.021 & .017 & .005 & .007 & .002 & \cellcolor{yellow!75}.041 & .003 & \cellcolor{gray!50}\textbf{.695} & .010 & .012 & .011 & .002 & \cellcolor{red!75}.108 & .001 & .008 \\
    \hline
    ru & .001 & .011 & .002 & .002 & .000 & .001 & .008 & .005 & .002 & .006 & .006 & .010 & .004 & \cellcolor{yellow!75}.039 & .004 & .009 & .002 & .007 & \cellcolor{gray!50}\textbf{.783} & \cellcolor{orange!75}.054 & .013 & .013 & .015 & .001 & .003 \\
    \hline
    sa & .001 & .006 & .002 & .002 & .000 & .000 & .006 & .003 & .002 & .006 & .015 & .016 & .002 & \cellcolor{orange!75}.052 & .001 & .009 & .001 & .011 & .013 & \cellcolor{gray!50}\textbf{.787} & .011 & .013 & \cellcolor{yellow!75}.032 & .001 & .006 \\
    \hline
    th & .001 & .008 & .002 & .003 & .000 & .001 & .007 & .004 & .003 & .008 & \cellcolor{orange!75}.057 & .015 & .003 & \cellcolor{yellow!75}.032 & .001 & .013 & .001 & .012 & .017 & \cellcolor{yellow!75}.038 & \cellcolor{gray!50}\textbf{.725} & .003 & \cellcolor{yellow!75}.041 & .001 & .004 \\
    \hline
    tr & .000 & .006 & .002 & .001 & .000 & .000 & .005 & .002 & .002 & .003 & .004 & .007 & .002 & .006 & .001 & .004 & .002 & .003 & .007 & .015 & .002 & \cellcolor{gray!50}\textbf{.917} & .007 & .000 & .003 \\
    \hline
    us & .003 & .014 & .005 & .015 & .001 & .003 & .007 & .006 & .004 & \cellcolor{yellow!75}.025 & .011 & .014 & .004 & .006 & .008 & .007 & .002 & .012 & .005 & .014 & .007 & .002 & \cellcolor{gray!50}\textbf{.811} & .003 & .008 \\
    \hline
    ve & \cellcolor{orange!75}.065 & .006 & .004 & .003 & .019 & .040 & .009 & \cellcolor{orange!75}.074 & .004 & .005 & .004 & .006 & .005 & .005 & .041 & .003 & .002 & .004 & .004 & .018 & .001 & .001 & .016 & \cellcolor{gray!50}\textbf{.659} & .002 \\
    \hline
    za & .001 & \cellcolor{yellow!75}.020 & .006 & .013 & .000 & .001 & .011 & .008 & .007 & \cellcolor{orange!75}.051 & .014 & \cellcolor{yellow!75}.032 & .005 & .006 & .001 & .012 & .011 & .016 & .006 & \cellcolor{yellow!75}.023 & .005 & .004 & \cellcolor{red!75}.175 & .001 & \cellcolor{gray!50}\textbf{.569} \\
    \hline
   \end{tabular}
   \caption{Aggregated confusion matrix for all classifiers on the top 25 countries. (ar: Argentina, au: Australia, br: Brazil, ca: Canada, cl: Chile, co: Colombia, de: Germany, es: Spain, fr: France, gb: United Kingdom, id: Indonesia, in: India, it: Italy, jp: Japan, mx: Mexico, my: Malaysia, nl: The Netherlands, ph: Philippines, ru: Russia, sa: Saudi Arabia, th: Thailand, tr: Turkey, us: United States, ve: Venezuela, za: South Africa)}
   \label{tab:confusion-matrix}
  \end{sidewaystable*}

On the positive side, some of the countries have very small misclassifications. Brazil and Turkey have misclassifications of less than 2\% (no yellow, orange or red cells). Other countries, including France, Indonesia, Italy, Japan and the USA, have misclassifications of less than 5\% (no red or orange cells). These are mostly countries with unique characteristics with respect to the rest of the top 25 countries; they predominantly use a language that is not used by any other in the list, except the USA, which has the advantage of having the majority of tweets. However, a striking observation is the large percentage of misclassifications involving Spanish speaking countries, which include Argentina, Chile, Colombia, Spain, Mexico and Venezuela. In most of these cases the high number of misclassifications occurs in both directions for each pair of countries. This is an additional difficulty that one might have expected, given that all of them share cultural and linguistic commonalities, especially for using the same language and hence overlapping content. Moreover, the Latin American countries often share the time zone and, while the time zone is different for Spain, many of the cities in the Latin American countries are named after Spanish cities (e.g., C\'ordoba in Argentina, Le\'on in Mexico, Valencia in Venezuela, Cartagena in Colombia or Santiago in Chile, all of which are also Spanish cities), which makes the distinction from Spain more challenging if only \textit{user location} is used. Similarly, we also observe a large amount of misclassifications involving English speaking countries, e.g. Australia, the UK, Canada and the USA. The majority of the orange misclassifications (5\%-10\%) are between Spanish and English speaking countries, with the exception of Chile and Argentina, which are even higher (10\%+) and which we surmise is due to their proximity and cultural similarities. Finally, many misclassifications involve the United States, which account for the majority of red misclassifications (10\%+), and which is not surprising since it is the predominant country with about 20\% of tweets.

\section{Discussion}

Our experiments and analysis on over 5 million geolocated tweets from unique users reveal insights into country-level geolocation of tweets in real time. Our experiments only make use of features inherent in the tweets to enable real-time classification. This can be invaluable when curation of the tweet stream is needed for applications such as country-specific trending topic detection \cite{atefeh2015survey}, or for more specific applications where only tweets coming from a specific country are sought, e.g. sentiment analysis or reputation management \cite{amigo2013overview}. The identification of the country of origin will also help mitigate problems caused by the limited availability of demographic details for Twitter users \cite{murthy2016urban}.

We found that one of the most commonly used approaches, which is the use of gazeteers such as GeoNames to match the user's self-reported location with a place in the world, performs reasonably well in terms of macro-accuracy, but fails in terms of micro-accuracy, i.e. without high accuracy for most countries. The use of a classifier that makes use of a single feature, such as the self-reported location of a user, outperforms the GeoNames baseline in terms of micro-accuracy, as well as slightly in terms of macro-accuracy. The main challenge is that it has to deal with as many as 217 countries, making the task especially difficult. To overcome this, we have tested our classifier on a reduced subset of the top 25 countries, which still account for more than 90\% of the whole Twitter stream. In this case, we found that this classifier can substantially outperform both the GeoNames baseline and the state-of-the-art real-time tweet geolocation classifier by \cite{dredze2016geolocation}. The use of the tweet content alone becomes then the most useful feature.

Further testing with combinations of multiple features, we found that performance can be substantially improved, although one needs to be careful when picking the features to be used. What is interesting is that the classifier trained on data from the same time frame as the test set can be effectively applied to new tweets, which we verified on tweets posted a year later. The combination of features that works well for the test set in the same time frame can be applied to the new tweets in most cases, achieving similar performance values. However, it is important to consider that the utility of some features drops over time, which is especially the case of \textit{user description}, but also to a lesser extent other features like \textit{offset} and \textit{tweet language}. On the positive side, features like \textit{tweet content}, \textit{user location} and \textit{user name} are among the most useful features for classifying new tweets. One may also choose to regularly update the classifier by training with new tweets, as \cite{dredze2016geolocation} suggested, however, in the interest of keeping a model for longer and reducing the cost of updating models, we show that the choice of the appropriate features can be as effective (i.e. achieving macro-accuracy scores of 0.858 and 0.853 for tweets within the same time frame and new tweets, respectively). The scenario is quite different when one wants to identify tweets from a specific country, given that different sets of features lead to more accurate classifications for different countries, which do not necessarily match with the overall best approach. By picking the right combination of features one can achieve classification performances for a country higher than 0.8 and even above 0.9 in terms of F1 score in cases where a country has unique characteristics such as a language that is not spoken in other countries or a unique time zone. However, these performance values tend to drop when one aims to identify tweets for a country that has common characteristics with other countries; this is especially true for English and Spanish speaking countries, among which many are large countries that speak the same language, share similar contents and have the same time zone (e.g., Chile and Argentina, or Canada and the USA).

The use of geolocated tweets to build a collection of tweets with a location assigned is a widely accepted practice, although the applicability of a model trained on geolocated tweets to then classify non-geolocated tweets has not been studied in depth. In previous work, \cite{han2014text} suggested that a model trained on geotagged data is expected to generalise well to non-geotagged data when one wants to classify users. For our case study with tweets rather than users, we performed a comparative analysis of geolocated and non-geolocated tweets in the time frame of our TC2014 dataset\footnote{Tweets were retrieved from the Internet Archive: \url{https://archive.org/details/archiveteam-twitter-stream-2014-10}}. Looking at the ranked frequencies for each feature, we found high correlations ranging from $r = 0.858$ to $r = 0.956$ for seven of the features under study across the subsets of geolocated and non-geolocated tweets, except for \textit{content} leading to lower correlation ($r = 0.295$). This indicates that non-geolocated tweets have similar characteristics and that a model trained on geolocated tweets could be effectively applied, reinforcing our findings that the use of content alone, as in most previous work, does not suffice, and combination of features is recommended. Empirical experimentation on non-geolocated tweets would help quantify this further; however an alternative data collection and annotation methodology should be defined for this purpose, which is beyond the scope of this work.

In summary, the results suggest that an appropriate selection of tweet features can lead to accurate, real-time classification of the most populous countries in terms of volume. Interestingly, a model trained from historical tweets can also be applied to tweets collected later in time when the topics that users talk about may be completely different. Having this classifier in place, one may then want to perform finer-grained geolocation of tweets within a country. For instance, during breaking news, one may want to identify reports from eyewitnesses on the ground and therefore fine-grained geolocation would be crucial to identify tweets in the area.

\section{Conclusion}

To the best of our knowledge, this is the first study performing a comprehensive analysis of the usefulness of tweet-inherent features to automatically infer the country of origin of tweets in a real-time scenario from a global stream of tweets written in any language. Most previous work focused on classifying tweets coming from a single country and hence assumed that tweets from that country were already identified. Where previous work had considered tweets from all over the world, the set of features employed for the classification included features, such as a user's social network, that are not readily available within a tweet and so is not feasible in a scenario where tweets need to be classified in real-time as they are collected from the streaming API. Moreover, previous attempts to geolocate global tweets tended to restrict their collection to tweets from a list of cities, as well as to tweets in English; this means that they did not consider the entire stream, but only a set of cities, which assumes prior preprocessing. Finally, our study uses two datasets collected a year apart from each other, to test the ability to classify new tweets with a classifier trained on older tweets. Our experiments and analysis reveal insights that can be used effectively to build an application that classifies tweets by country in real time, either when the goal is to organise content by country or when one wants to identify all the content posted from a specific country.

In the future we plan to test alternative cost-sensitive learning approaches to the one used here, focusing especially on collection of more data for under-represented countries, so that the classifier can be further improved for all the countries. Furthermore, we plan to explore more sophisticated approaches for content analysis, e.g. detection of topics in content (e.g. do some countries talk more about football than others?), as well as semantic treatment of the content. We also aim to develop finer-grained classifiers that take the output of the country-level classifier as input.

\section*{Acknowledgments}

This work has been supported by the PHEME FP7 project (grant No. 611233), the Warwick University Higher Education Impact Fund, an ESRC Impact Acceleration Award, EPSRC Impact Acceleration Account (grant no. EP/K503940/1) and EPSRC grant EP/L016400/1. We used the MidPlus computational facilities, supported by QMUL Research-IT and funded by EPSRC grant EP/K000128/1.

\end{document}